\documentclass[pre,reprint,superscriptaddress]{revtex4-1}

\usepackage{amsmath,amssymb}

\usepackage{mathtools}

\usepackage{graphicx}

\usepackage{hyperref}

\newcommand{\beq}{\begin{equation}}
\newcommand{\eeq}{\end{equation}}

\begin{document}

\title{Astrocyte-induced  positive integrated information in neuroglial 
ensembles}

\author{Oleg Kanakov}
\author{Susanna Gordleeva}
\author{Anastasia Ermolaeva}
\affiliation{Lobachevsky State University of Nizhny Novgorod, Nizhny Novgorod, 
Russia}	

\author{Sarika Jalan}
\affiliation{Complex Systems Lab, Discipline of Physics, Indian Institute of 
Technology Indore, Simrol, Indore-452020}	

\author{Alexey Zaikin}\email[E-mail: ]{alexey.zaikin@ucl.ac.uk}
\affiliation{Lobachevsky State University of Nizhny Novgorod, Nizhny Novgorod, 
Russia}	
\affiliation{Institute for Women's Health and Department of Mathematics, 
University College London, London, United Kingdom}

\begin{abstract}
The Integrated Information is a quantitative measure from information theory 
how tightly all parts of a system are interconnected in terms of information 
exchange. In this study we show that astrocyte, playing an important role in 
regulation of information transmission between neurons, may contribute to a 
generation of positive Integrated Information in neuronal ensembles. 
Analytically and numerically we show that the presence of astrocyte may be 
essential for this information attribute in neuro-astrocytic ensembles. 
Moreover, the proposed ``spiking-bursting'' mechanism of generating positive 
Integrated Information is shown to be generic and not limited to neuroglial 
networks, and is given a complete analytic description.
\end{abstract}

\maketitle

\section{Introduction}

The Integrated Information (\emph{II}) concept introduced
in \cite{tononi2004information} marked a milestone in the ongoing effort to
describe activities of neural ensembles and brain by means of
information theory. \emph{II} was proposed as a quantitative
measure of how tightly all parts of a system are interconnected in
terms of information exchange (for example, a combination of two
non-interacting subsystems implies zero \emph{II}). The ambitious
aim of the \emph{II} concept was to quantify consciousness 
\cite{tononi2008consciousness}
--- in particular, for medical applications to detecting
consciousness in a completely immobilized patient by
electroencephalographic data. Several mathematical definitions of
\emph{II} \cite{barrett2011practical,tononi2012integrated, 
oizumi2014phenomenology, tegmark2016improved}  have been proposed since the 
original work,
all in line with the initial idea. The perturbational complexity index, linked 
to \emph{II} as its proxy, has reliably discriminated the level of 
consciousness in patients during wakefulness, sleep, anesthesia, and even in 
patients who has emerged from coma with minimal level of consciousness 
\cite{Casali198ra105}. Although the relation of
\emph{II} to consciousness has been debated \cite{peressini2013consciousness, 
tsuchiya2016using,2017arXiv170802967T}, \emph{II}
itself is by now widely adopted
as a quantitative measure for complex dynamics  \cite{tononi2016integrated, 
engel2017integrated, norman2017quantum}. Accordingly, the
understanding of particular mechanisms producing positive
\emph{II} in neural ensembles is of topical interest.

The experiments have shown that astrocytes play an important role by regulating 
cellular functions and information transmission in the nervous system 
\cite{perea2010, araque2014}. It was  proposed that astrocyte wrapping a 
synapse implements a feedback control circuit which maximizes information 
transmission through the synapse by regulating neurotransmitter release 
\cite{nadkarni2008astrocytes}. The involvement of astrocytes in neuro-glial 
network dynamics was quantified by estimating functional connectivity between 
neurons and astrocytes from time-lapse Ca$^{2+}$ imaging data 
\cite{nakae2014statistical}.
In contrast with neuronal cells the astrocytes do not generate electrical 
excitations (action potentials). However, their intracellular dynamics have 
shown similar excitable properties for changes of calcium concentration 
\cite{nadkarni2003}. These signals can remarkably affect neuronal excitability 
and the efficiency of synaptic transmission between neurons by 
Ca$^{2+}$-dependent release of neuroactive chemicals (e.g. glutamate, ATP, 
D-serine and GABA) \cite{parpura2010gliotransmission}. 
Networks of astrocytes accompanying neuronal cells generate collective activity 
patterns that can regulate neuronal signaling by facilitating or by suppressing 
synaptic transmission \cite{perea2010, araque2014, depitta2016}. 

In this study we show that astrocytes may conduce to positive \emph{II} in 
neuronal ensembles. We calculate \emph{II} in a small neuro-astrocytic network 
with random topology by numerical simulation and find that positive \emph{II} 
is conditioned by coupling of neurons to astrocytes and increases with 
spontaneous neuronal spiking activity. We explain this behavior using 
simplified spiking-bursting dynamics, which we implement both in the 
neuro-astrocytic network model with all-to-all connectivity between neurons, 
showing astrocyte-induced coordinated bursting, and as well in a specially 
defined stochastic process allowing analytical calculation of \emph{II}. The 
analytical and simulation results for the all-to-all network are in good 
agreement. That said, non-trivial dynamics of the random version of the 
network, although not being directly compatible with our analytical treatment, 
turns out to be even more favorable for positive \textit{II} than the 
spiking-bursting dynamics of the all-to-all network. We speculate that the 
presence of astrocytes may be essential for generating positive \emph{II} in 
larger neuro-astrocytic ensembles.

\section{Methods and Model}\label{sec_meth_mod}
\begin{figure*} %[h!]
	\centering
	A\includegraphics[width=0.95\columnwidth]{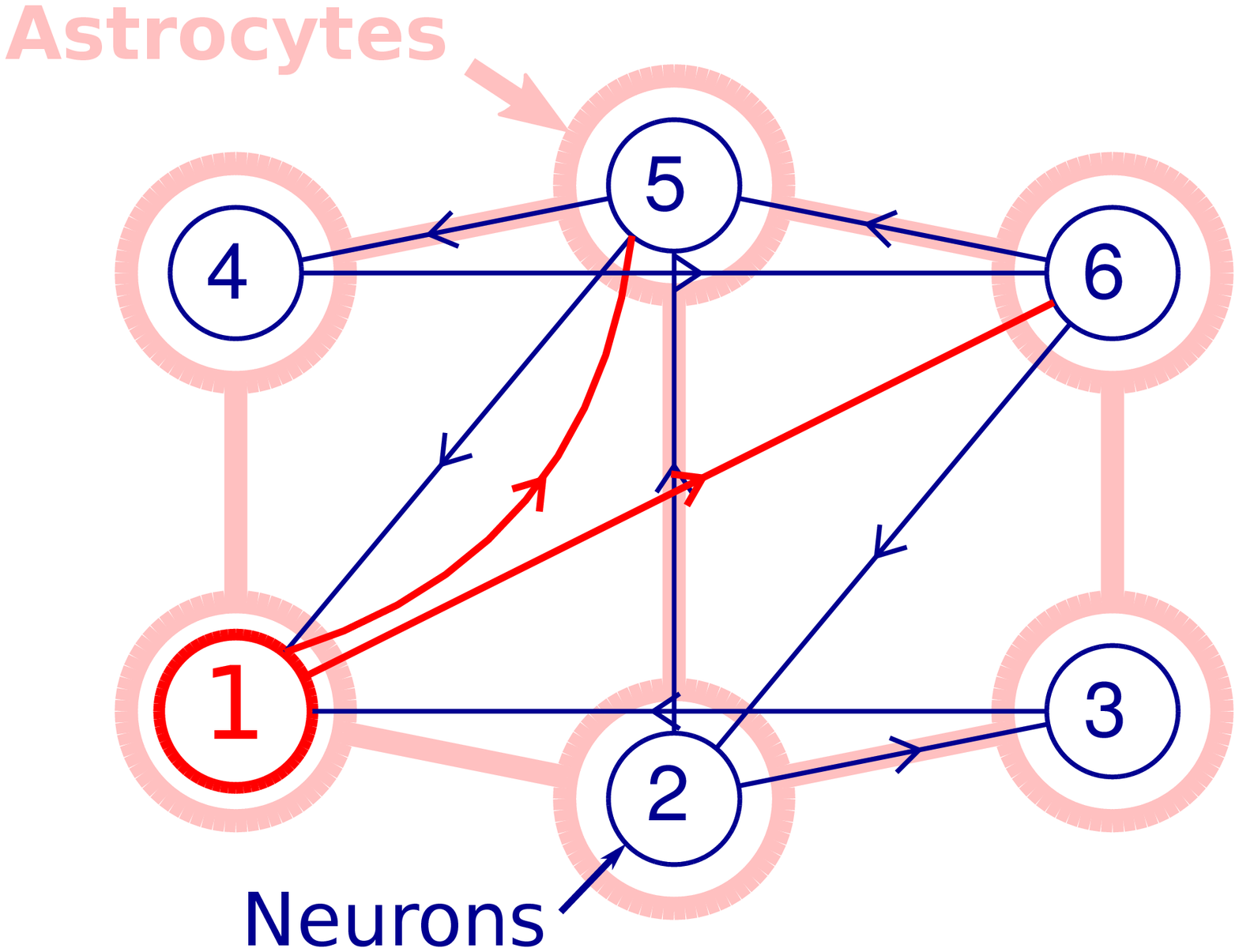}
	B\includegraphics[width=0.95\columnwidth]{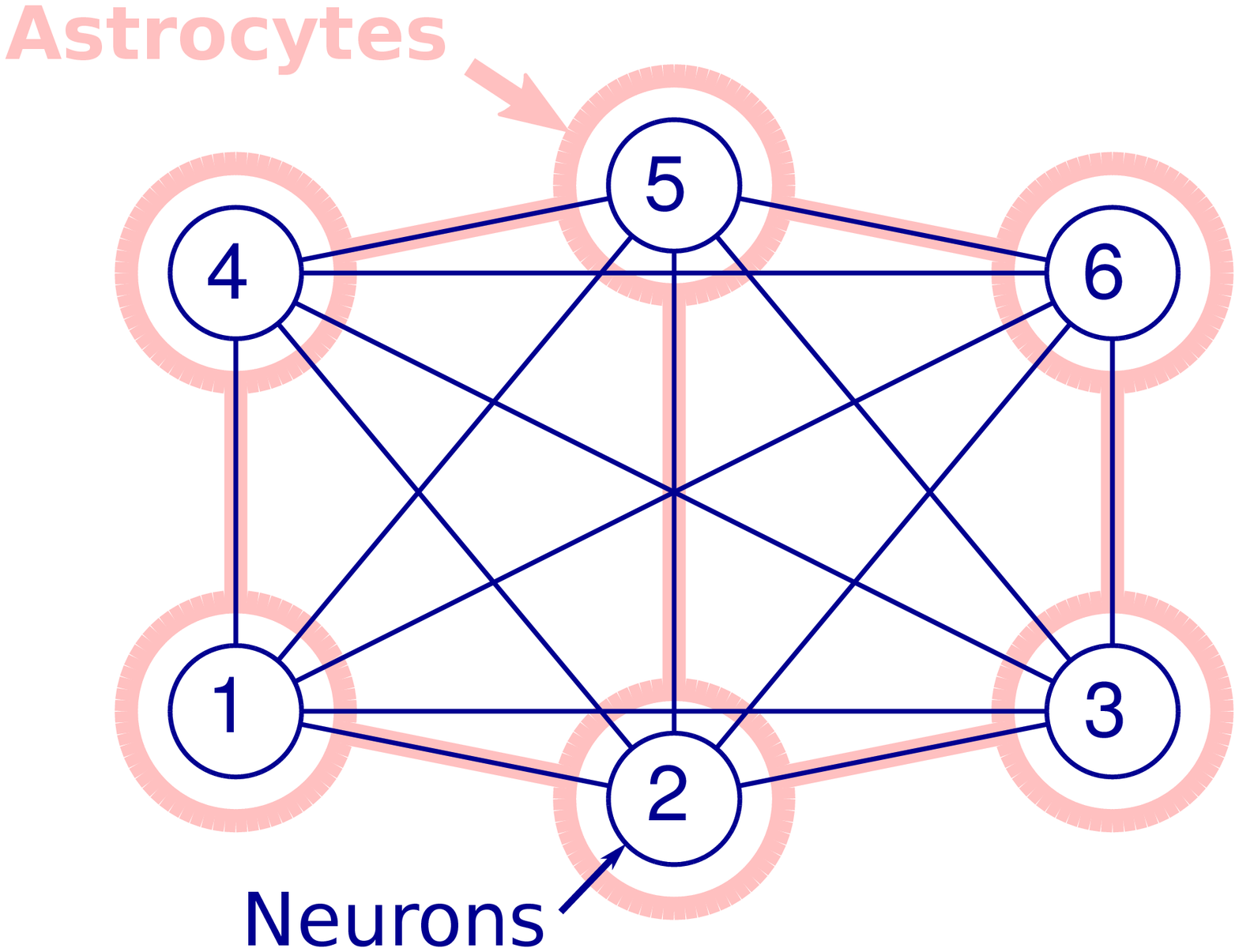}
			\caption{(Color online) Schemes of neuro-astrocytic 
networks under study: A --- instance of random neuronal network topology; B --- 
all-to-all neuronal network. Inhibitory neuron is shown with an enlarged symbol 
and highlighted in red. Connections without arrows are bi-directional. Each 
astrocyte is coupled to one corresponding neuron and acts by modulating 
outgoing connections of the neuron.}\label{fig_topo}
\end{figure*}

Neural network under study consists of 6 synaptically coupled Hodgkin-Huxley 
neurons \cite{hodgkin1952}. We use 2 variants of neural network architecture: 
(i) network of 1 inhibitory and 5 excitatory neurons with coupling topology 
obtained by randomly picking 1/3 of the total number of connections out of the 
full directed graph, excluding self-connections (the particular instance of 
random topology for which the presented data have been obtained is shown in 
Fig.~\ref{fig_topo}A); (ii) all-to-all network of 6 excitatory neurons 
(Fig.~\ref{fig_topo}B).

The membrane potential of a single neuron evolves according to the following 
ionic current balance equation:
\begin{equation}
\label{eq:HH_main}
C \frac{dV^{(i)}}{dt} = I^{(i)}_{\text{channel}} + I^{(i)}_{\text{app}} + 
\sum_j {I^{(ij)}_{\text{syn}}} + I^{(i)}_{P};
\end{equation}
where the superscript ($i = 1,\ldots, 6$) corresponds to a neuronal index and 
($j$) corresponds to an index of input connection. Ionic currents (i.e. sodium, 
potassium and leak currents) are expressed as follows:
\begin{equation}
\label{eq:HH_currents}
\begin{aligned}
&\begin{multlined}
    I_{\text{channel}} = -g_{Na}m^{3}h(V-E_{Na})-{}\\
    -g_Kn^{4}(V-E_K)-g_{\text{leak}}(V-E_{\text{leak}}),
\end{multlined}\\
   & \frac{dx}{dt} = \alpha_{x}(1-x)-\beta_{x}x, \;\;\;  x=m,n,h.
\end{aligned}
\end{equation}
Nonlinear functions $\alpha_x$ and $\beta_x$ for gating variables are taken 
as in original Hodgkin-Huxley model with membrane potential $V$ shifted by 65 
mV. Throughout this paper we use the 
following parameter values: $E_{Na} = 55$ mV, $E_{K} = -77$ mV, 
$E_{\text{leak}} = -54.4$ mV, $g_{Na} = 120$ mS/cm$^2$, $g_{K} = 36$ mS/cm$^2$, 
$g_{\text{leak}} = 0.3$ mS/cm$^2$, $C = 1$ $\mu$F/cm$^2$. The 
applied currents 
$I^{(i)}_{\text{app}}$ are fixed at constant value controlling the 
depolarization level and dynamical regime that can be either excitable, 
oscillatory or bistable \cite{kazantsev2011}. We use 
$I^{(i)}_{\text{app}}=-5.0$ $\mu$A/cm\textsuperscript{2} which corresponds to 
excitable regime. The synaptic current 
$I_{\text{syn}}$ simulating interactions between the neurons obeys the equation:
\begin{equation}
    \label{eq:I_syn}
    \begin{aligned}
    &I^{(ij)}_{\text{syn}} = 
\frac{g_{\text{syneff}}(V^{(j)}-E_{\text{syn}})}{1+\exp(\frac{-(V^{(i)}-\theta_{
\text{syn}})}{k_{\text{syn}}})};\\
    \end{aligned}
\end{equation}
where $E_{\text{syn}}=-90$ mV for the inhibitory synapse and $E_{\text{syn}}=0$ 
mV for the excitatory. Neural network composition of 1 inhibitory and 5 
excitatory neurons is in line with the experimental data showing that the 
fraction of inhibitory neurons is about 20\% \cite{braitenberg1991}. Variable 
$g_{\text{syneff}}$ describes the synaptic weight in mS/cm$^2$ modulated by an 
astrocyte (as defined by Eq.~\eqref{eq:astro_neuro} below), parameters 
$\theta_{\text{syn}}=0$ mV and $k_{\text{syn}}=0.2$ mV describe the midpoint of 
the synaptic activation and the slope of its threshold, respectively. 

Each neuron is stimulated by a Poisson pulse train mimicking external spiking 
inputs $I^{(i)}_{P}$ with a certain average rate $\lambda$. Each Poisson pulse 
has constant duration 10 ms and constant amplitude, which is sampled 
independently for each pulse from uniform random distribution on interval 
$[-1.8, 1.8]$. Sequences of Poisson pulses applied to different neurons are 
independent.

Note that the time unit in the neuronal model \eqref{eq:HH_main}, 
\eqref{eq:HH_currents} is one millisecond. Due to a slower time scale, in the 
astrocytic model (see below) empirical constants are indicated using seconds as 
time units. When integrating the joint system of differential equations, the 
astrocytic model time is rescaled so that the units in both models match up.

 We consider astrocytic 
network in the form of a two-dimensional square lattice with only 
nearest-neighbor connections \cite{kazantsev2009}. Such topology for the 
Ca$^{2+}$- and IP$_3$-diffusion model is justified by experimental findings 
stating that astrocytes occupy ``nonoverlapping'' territories 
\cite{halassa2007}. The neuro-astrocyte network of real brain has a 3D 
structure with one astrocyte interacting with several neurons and vise versa. 
However, in our modelling we use a simplified approach. The latter reflects the 
fact that, throughout area CA1 of the 
hippocampus, pyramidal (excitatory) cells are arranged in a regular layer and 
surrounded by a relatively uniform scatter of astrocytes \cite{ferrante2015}. 
Accordingly to the experimental data \cite{savtchenko2014, ferrante2015}, 
modelled astrocytes are distributed evenly across the neural network, with a 
total cell number equaled to the number of neurons (due to small size of 
networks, astrocyte network is modeled as a 2-D lattice in our case study). 
Astrocytes and neurons communicate via a special mechanism modulated by 
neurotransmitters from both sides. The model is designed so that when the 
calcium level inside an astrocyte exceeds a threshold, the astrocyte releases 
neuromodulator (here, glutamate) that may affect the 
release probability (and thus a synaptic strength) at neighboring connections 
in a tissue volume \cite{navarrete2010endocannabinoids}. Single astrocyte can 
regulate the synaptic strength of several neighboring synapses which belong to 
one neuron or several different neurons, but since we do not take into account 
the complex morphological structure of the astrocyte, we assume for simplicity 
that one astrocyte interacts with one neuron.

In a number of previous studies a biophysical mechanism underlying calcium 
dynamics of astrocytes has been extensively investigated \cite{deyoung1992, 
ullah2006}. Calcium is released from internal stores, mostly from the 
endoplasmic reticulum (ER). This process is regulated by inositol 
1,4,5-trisphosphate (IP$_3$) that activates IP$_3$ channels in the ER membrane 
resulting in a Ca$^{2+}$ influx from ER. IP$_3$ acting as a second messenger is 
produced when neurotransmitter (i.e. glutamate) molecules are bound by 
metabotropic receptors of the astrocyte. In turn IP$_3$ can be regenerated 
depending on the level of calcium by the phospholipase C-$\delta$ 
(PLC-$\delta$). State variables of each cell include IP$_3$ concentration 
$IP_3$, Ca$^{2+}$ concentration $Ca$, and the fraction of activated IP$_3$ 
receptors $h$. They evolve according to the following equations 
\cite{deyoung1992, ullah2006}:
\begin{flalign}
\label{eq:astro_main}
&\begin{aligned}
&\begin{multlined}[t][0.8\displaywidth]
\frac{dCa^{(m,n)}}{dt} =  
J_{\text{ER}}^{(m,n)}-J_{\text{pump}}^{(m,n)}+J_{\text{leak}}^{(m,n)}+{}\\
+J_{\text{in}}^{(m,n)}-J_{\text{out}}^{(m,n)}
+J^{(m,n)}_{Ca\text{diff}};
\end{multlined}\\
&\mathrlap{\frac{dIP_3^{(m,n)}}{dt} =  
\frac{IP_3^*-IP_3^{(m,n)}}{\tau_{IP3}}+J_{\text{PLC}}^{(m,n)} %+{}\\
+J^{(m,n)}_{IP3\text{diff}};
}
\\
&\mathrlap{
	\begin{multlined}[t][0.8\displaywidth]
\frac{dh^{(m,n)}}{dt} = \\ 
=a_2\left(d_2\frac{IP_3^{(m,n)}+d_1}{IP_3^{(m,n)}+d_3}(1-h^{m,n})-Ca^{m,n}h^{m,n
}\right);
\end{multlined}}
\end{aligned}&
\end{flalign}
with $m=1,\ldots,3$, $n=1,2$. Currents $J_{\text{ER}}$ is Ca$^{2+}$ current 
from the ER to the cytoplasm, $J_{\text{pump}}$ is the ATP pumping current, 
$J_{\text{leak}}$ is the leak current, $J_{\text{in}}$ and $J_{\text{out}}$ 
describe calcium exchanges with extracellular space, $J_{\text{PLC}}$ is the 
calcium-dependent PLC-$\delta$ current and are expressed as follows:
\begin{equation}
\label{eq:astro_currents}
\begin{aligned}
&J_{\text{ER}} = 
c_1v_1Ca^3h^3IP_3^3\frac{(c_0/c_1-(1+1/c_1)Ca)}{((IP_3+d_1)(IP_3+d_5))^3};\\
&J_{\text{pump}} = \frac{v_3Ca^2}{k_3^2+Ca^2};\\
&J_{\text{leak}} = c_1v_2(c_0/c_1-(1+1/c_1)Ca);\\
&J_{\text{in}} = v_5+\frac{v_6 IP_3^2}{k_2^2+IP_3^2};\\
&J_{\text{out}} = k_1Ca;\\
&J_{\text{PLC}} = \frac{v_4(Ca+(1-\alpha)k_4)}{Ca+k_4}.
\end{aligned}
\end{equation}
Biophysical meaning of all parameters in Eqs. \eqref{eq:astro_main}, 
\eqref{eq:astro_currents} and their values determined experimentally can be 
found in \cite{deyoung1992, ullah2006}. For our purpose we fix $c_0 = 2.0$ 
$\mu$M, $c_1 = 0.185$, $v_1 = 6$ s$^{-1}$, $v_2 = 0.11$ s$^{-1}$, $v_3 = 2.2$ 
$\mu$Ms$^{-1}$, $v_5 = 0.025$ $\mu$Ms$^{-1}$, $v_6 = 0.2$ $\mu$Ms$^{-1}$, $k_1= 
0.5$ s$^{-1}$, $k_2 = 1.0$ $\mu$M, $k_3 = 0.1$ $\mu$M, $a_2 = 0.14$ 
$\mu$M$^{-1}$s$^{-1}$, $d_1 = 0.13$ $\mu$M, $d_2 = 1.049$ $\mu$M, $d_3 = 
0.9434$ $\mu$M, $d_5 = 0.082$ $\mu$M, $\alpha = 0.8$, $\tau_{IP3} = 7.143$ s, 
$IP_3^*=  0.16$  $\mu$M, $k_4 = 1.1$ $\mu$M \footnote{For aligning the 
time units of the neuronal and astrocytic parts of the model it is sufficient 
to re-express the numerical values of all 
dimensional constants using time unit of 1 ms}. Parameter $v_4$ 
describes the rate 
of $IP_3$ regeneration and controls the dynamical regime of the model 
\eqref{eq:astro_main}, \eqref{eq:astro_currents} that can be excitable at 
$v_4=0.3$ $\mu$Ms$^{-1}$, or oscillatory at $v_4=0.5$ $\mu$Ms$^{-1}$ 
\cite{ullah2006}. Here we limit ourselves to the oscillatory case.

Currents $J_{Ca\text{diff}}$ and $J_{IP3\text{diff}}$ describe the diffusion of 
Ca$^{2+}$ ions and IP$_3$ molecules via gap junctions between astrocytes in the 
network and can be expressed as follows \cite{kazantsev2009}:
\begin{equation}
\label{eq:astro_dif}
\begin{aligned}
&J^{(m,n)}_{Ca\text{diff}} = d_{\text{Ca}}(\Delta Ca)^{(m,n)};\\
&J^{(m,n)}_{IP3\text{diff}} = d_{\text{IP3}}(\Delta IP_3)^{(m,n)};\\
\end{aligned}
\end{equation}
where parameters $d_{\text{Ca}} = 0.001$ s$^{-1}$ and $d_{\text{IP3}} = 0.12$ 
s$^{-1}$ describe the Ca$^{2+}$ and IP$_3$ diffusion rates, respectively. 
$(\Delta Ca)^{(m,n)}$ and $(\Delta IP_3)^{(m,n)}$ are discrete Laplace 
operators:  
\begin{multline}
\label{eq:astro_Laplace}
(\Delta Ca)^{(m,n)} =(Ca^{(m+1,n)}+Ca^{(m-1,n)}\\+Ca^{(m,n+1)}
+Ca^{(m,n-1)}-4Ca^{(m,n)}).
\end{multline}

Astrocytes can modify release probability of nearby synapses in tissue volume 
\cite{perea2010}, likely by releasing signalling molecules ('gliotransmitters') 
in a Ca$^{2+}$ dependent manner \cite{araque2014}. We proposed that each 
astrocyte from the network interacts to the one neuron from the neural network 
by modulation of the synaptic weight. For the sake of simplicity, the effect of 
astrocyte calcium concentration $Ca$ upon synaptic weight of the affected 
synapses $g_{\text{syneff}}$ (which appears in Eq.~\eqref{eq:I_syn}) has been 
described with the simple formalism based on earlier suggestions 
\cite{volman2007, depitta2011, gordleeva2012}:
\begin{equation}
\label{eq:astro_neuro}
g_{\text{syneff}}=\begin{cases}
g_{\text{syn}}(1+g_{\text{astro}}Ca^{(m,n)}), & \text{if } Ca^{(m,n)} >0.2,\\
g_{\text{syn}}, & \text{otherwise}; % \text{if } Ca^{(m,n)} \le 0.2;
\end{cases}
\end{equation}    
where $g_{\text{syn}}=0.04$ mS/cm\textsuperscript{2} is baseline synaptic 
weight, parameter 
$g_{\text{astro}}$ controls the strength of synaptic weight modulation, and 
$Ca^{(m,n)}$ is the intracellular calcium concentration in the astrocyte 
\eqref{eq:astro_main}. We assume $g_{\text{astro}}>0$, in consistency with 
available experimental data for excitatory synapses 
\cite{jourdain2007glutamate}.

The time series of neuron membrane potentials $V^{(i)}(t)$ are converted into 
binary-valued discrete-time processes according to \cite{archer2013bayesian} as 
follows. Time is split into windows of duration $T$ which become units of the 
discrete time. If inequality $V^{(i)}(t)>V_{\text{thr}}=-40.0$ mV is satisfied 
for at least some $t$ within a particular time window (essentially, if there 
was a spike in this time window), than the corresponding binary value (bit) is 
assigned 1, and 0 otherwise. The size of time window is chosen so that 
spontaneous spiking activity produces time-uncorrelated spatial patterns, but a 
burst shows as a train of successive 1's in the corresponding bit.

We use the definition of \emph{II} according to
\cite{barrett2011practical} as follows. Consider a stationary
stochastic process $\xi(t)$ (binary vector process), whose
instantaneous state is described by $N=6$ bits. The
full set of $N$ bits (``system'') can be split into two
non-overlapping non-empty subsets of bits (``subsystems'') $A$ and
$B$, such splitting further referred to as bipartition $AB$.
Denote by $x=\xi(t)$ and $y=\xi(t+\tau)$ two states of the process
separated by a specified time interval $\tau\neq0$. States of the
subsystems are denoted as $x_A$, $x_B$, $y_A$, $y_B$.

Mutual information between $x$ and $y$ is defined as
\beq\label{eq_defIxy}
I_{xy}=H_x+H_y-H_{xy},
\eeq
where $H_x=-\sum_x p_x \log_2 p_x$ is
entropy (base 2 logarithm gives
result in bits), $H_y=H_x$ due to stationarity which is assumed. Next, a
bipartition $AB$ is considered, and ``effective information'' as a
function of the particular bipartition is defined as
\beq
\label{eq_Ieff}
I_{\text{eff}}(AB)=I_{xy}-I_{x_A,y_A}-I_{x_B,y_B}.
\eeq

\emph{II} is then defined as effective information
calculated for a specific bipartition $AB^{\text{MIB}}$ (``minimum
information bipartition'') which minimizes specifically normalized
effective information:
\begin{subequations}\label{eq_II}
	\begin{gather}
		\mathit{II}=I_{\text{eff}}(AB^{\text{MIB}}),\\		
AB^{\text{MIB}}=\mathrm{argmin}_{AB} 
\left[\frac{I_{\text{eff}}(AB)}{\min\{H(x_A),H(x_B)\}} \right].\label{eq_II_b}
	\end{gather}
\end{subequations}
Note that this definition prohibits positive \emph{II}, when $I_{\text{eff}}$ 
turns out to be zero or negative for at least one bipartition $AB$.

In essence, mutual information measures the degree of dependence between two 
random events. In case of causality, when dependence is unidirectional, one can 
speak of degree of predictability instead. In this sense, effective information 
\eqref{eq_Ieff} measures how much the system is more predictable as a whole 
than when trying to predict the subsystems separately. Obvious cases when 
$I_{\text{eff}}$ is zero are (i) independent subsystems (then system as a whole 
is equally predictable as a combination of the parts) and (ii) complete absence 
of predictability (when all mutual informations are zero). When the system is 
fully synchronized (all bits are equal in any instance of time), for any 
bipartition we get $I_{xy}=I_{x_A,y_A}=I_{x_B,y_B}$, which implies 
$I_{\text{eff}}<0$ according to \eqref{eq_Ieff}. From (\ref{eq_II}a,b) we 
conclude that \emph{II} is zero or negative in the mentioned cases.

\section{Results}

\begin{figure*} % [h!]
	A\includegraphics[width=0.95\columnwidth]{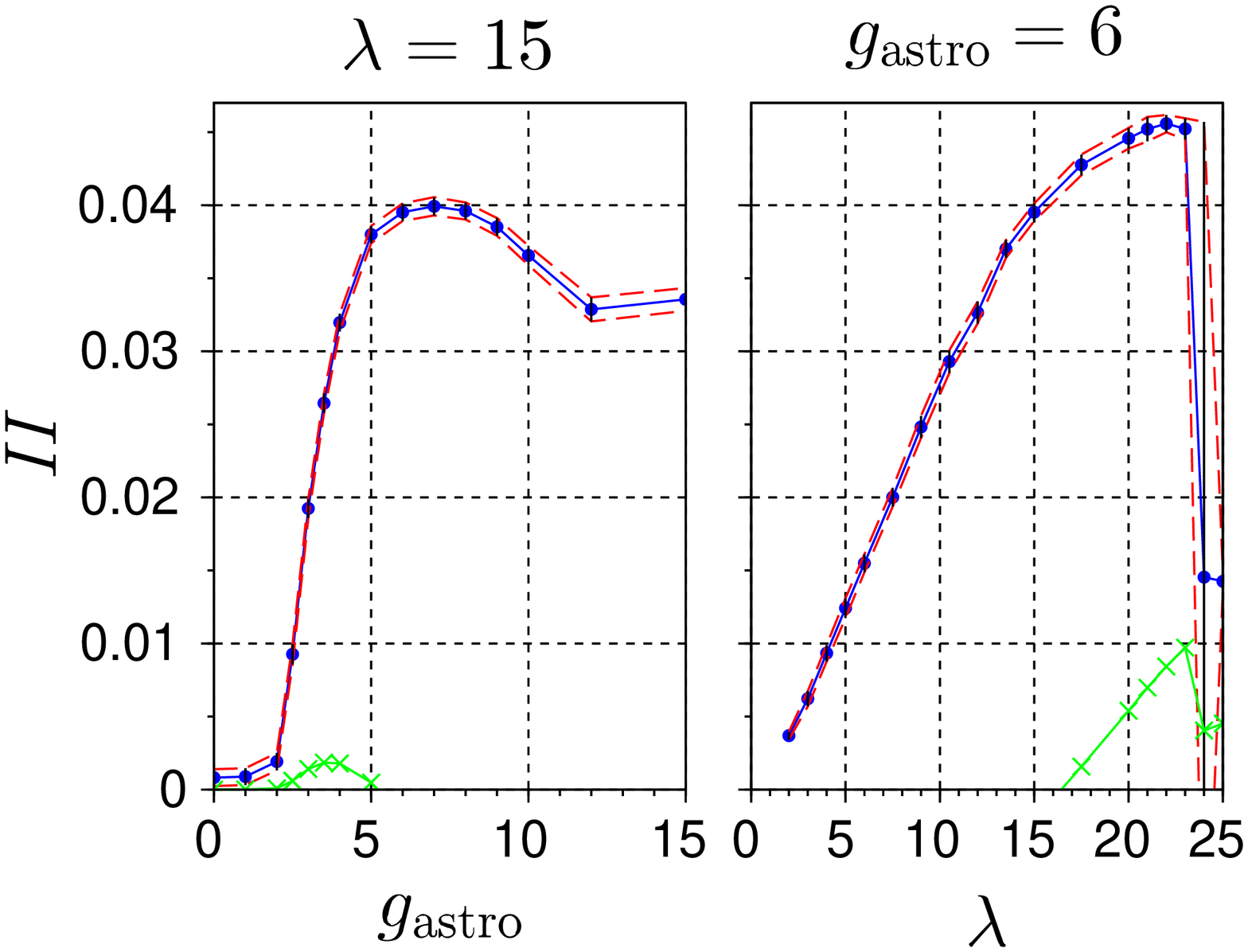}\hfill
	B\includegraphics[width=0.95\columnwidth]{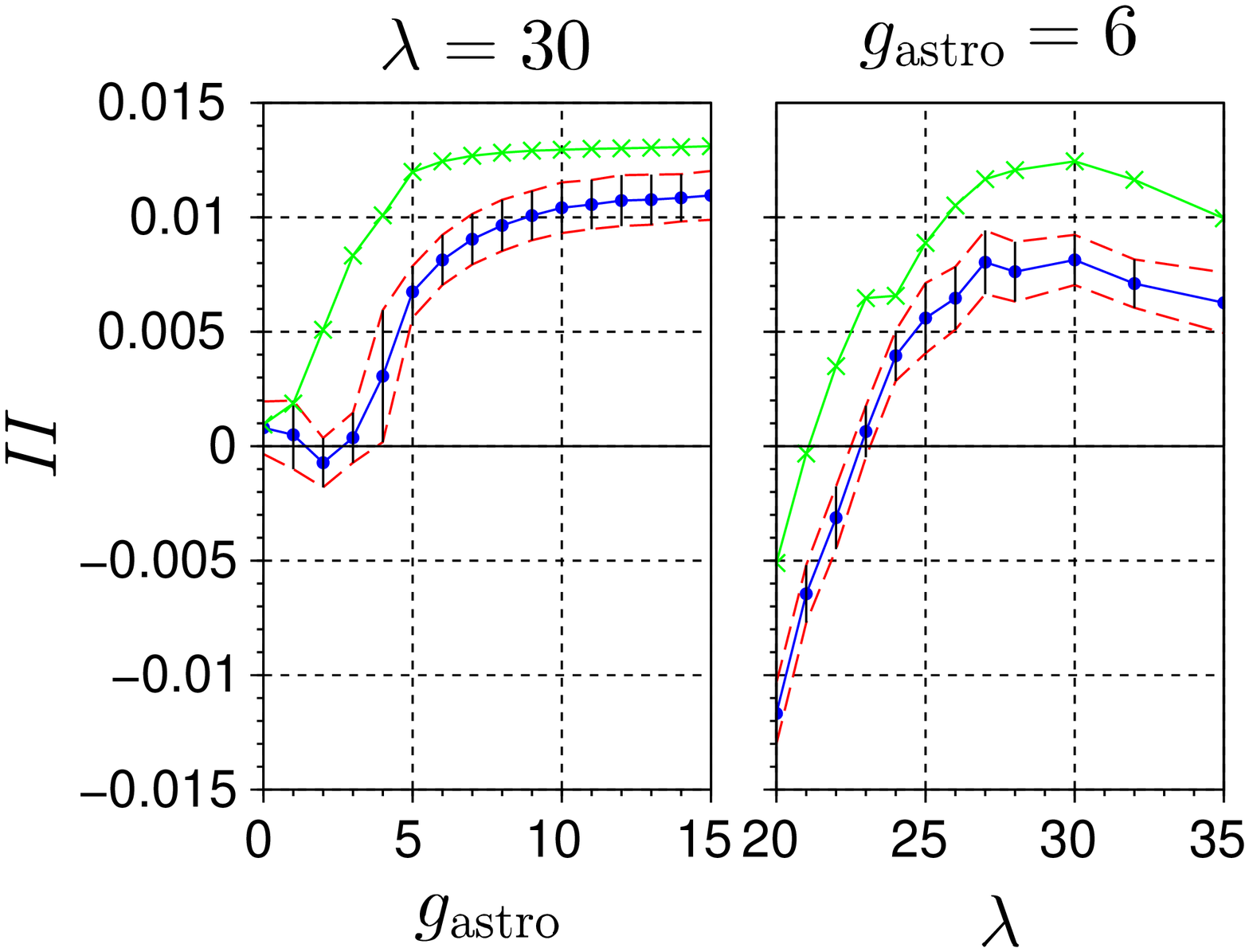}
	\caption{(Color online) Dependence of \emph{II} upon neuro-astrocytic 
interaction $g_{\text{astro}}$ and upon average stimulation rate $\lambda$: A 
--- in random network (instance shown in Fig.~\ref{fig_topo}A); B --- in 
all-to-all network (Fig.~\ref{fig_topo}B). Blue solid lines with dot marks --- 
direct calculation by definition from simulation data; red dashed lines --- 
error estimation; green lines with cross marks --- analytical calculation for 
spiking-bursting process with parameters estimated from simulation 
data.}\label{fig_graphs}
\end{figure*}

We calculated \textit{II} directly, according to definition above, using 
empirical probabilities from binarized time series of simulated 
neuro-astrocytic networks of both mentioned architectures Fig.~1A,B. For each 
architecture we performed two series of simulation runs: (i) with constant 
Poissonian stimulation rate $\lambda$ (equal 15.0~Hz for the random network and 
30.0~Hz for the all-to-all network) and neuro-astrocytic interaction 
$g_{\text{astro}}$ varied, (ii) with constant $g_{\text{astro}}=6.0$ and 
$\lambda$ varied, other model parameters as indicated above. Time window $T$ 
used in binarization and time delay $\tau$ used in computation of \emph{II} are 
$\tau=T=0.1$~s for the random network, and $\tau=T=0.2$~s for the all-to-all 
network. The length of time series to calculate each point is $5\cdot10^5$~s, 
taken after $2\cdot10^3$~s transient time. The estimate of \emph{II} shows 
convergence as the length of time series is increased. Error due to finite data 
(shown as half-height of errorbar in the graphs) is estimated as maximal 
absolute difference between the result for the whole observation time and for 
each its half taken separately.
Obtained dependencies of \textit{II} upon $g_{\text{astro}}$ and $\lambda$ are 
shown in Fig.~\ref{fig_graphs}.

For the random topology (Fig.~\ref{fig_graphs}A) we observe that (i) positive 
\emph{II} is greatly facilitated by non-zero $g_{\text{astro}}$ (i.e. by the 
effect of astrocytes), although small positive quantities, still exceeding the 
error 
estimate, are observed even at $g_{\text{astro}}=0$; (ii) \emph{II} 
generally increases with the average stimulation frequency $\lambda$ which 
determines the spontaneous activity in the network \footnote{The abrupt drop of 
\emph{II} at high $\lambda$ is associated with a change of minimum information 
bipartition and currently has no analytical explanation}.

\begin{figure*} % [h!]
	A\includegraphics[width=0.95\columnwidth]{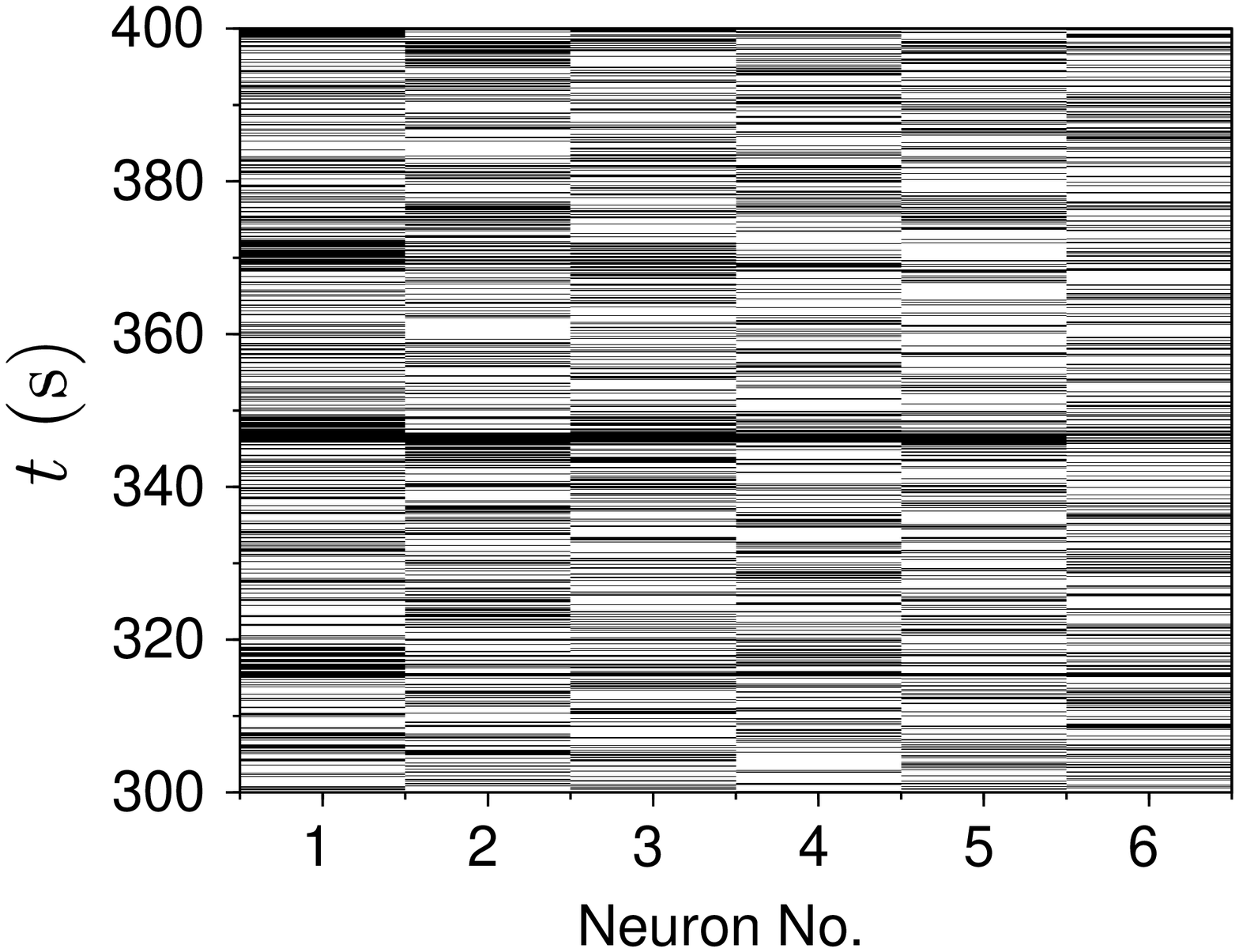}\hfill
	B\includegraphics[width=0.95\columnwidth]{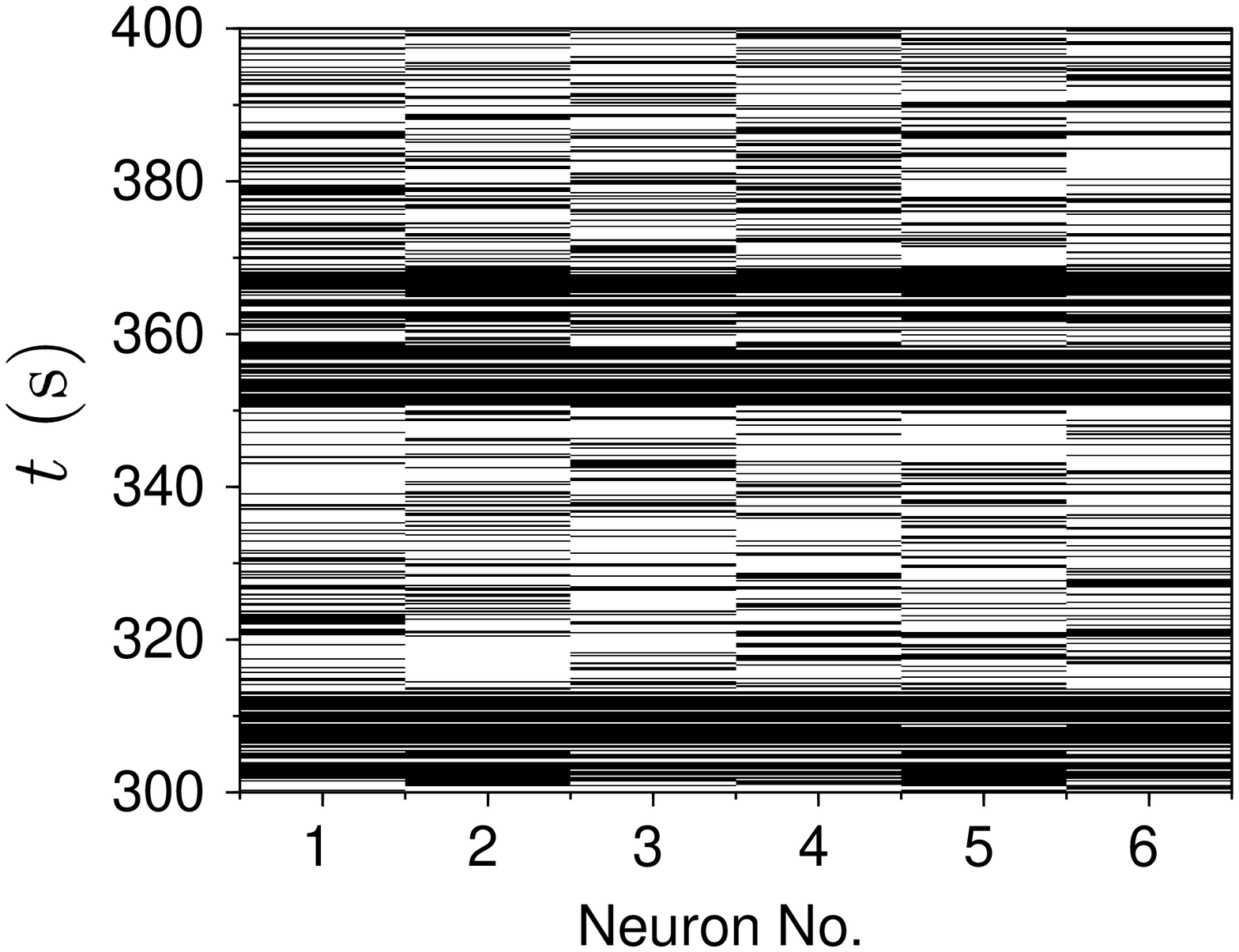}
	\caption{Raster plots of neuronal dynamics at $g_{\text{astro}}=6$, 
$\lambda=20$: A --- in random network (instance shown in Fig.~\ref{fig_topo}A); 
B --- in all-to-all network (Fig.~\ref{fig_topo}B). White and black correspond 
to 0 and 1 in binarized time series.}\label{fig_Raster}
\end{figure*}

The visible impact of astrocytes on the network dynamics consists in the 
stimulation of space-time patterns of neuronal activity due to 
astrocyte-controlled increase in the neuronal synaptic connectivity on 
astrocyte time scale. An instance of such pattern of activation for the random 
network is shown as a raster plot in Fig.~\ref{fig_Raster}A. The pattern is 
rather complex, and we only assume that \emph{II} must be determined by 
properties of this pattern, which in turn is controlled by astrocytic 
interaction. We currently do not identify specific properties of activation 
patterns linked to the behaviour of \emph{II} in the random network; however, 
we do it (see below) for the all-to-all network of identical (all excitatory) 
neurons, due to its simpler ``spiking-bursting'' type of spatio-temporal 
dynamics consisting of coordinated system-wide bursts overlaid upon
background spiking activity, see raster plot in Fig.~\ref{fig_Raster}B. As seen 
in  Fig.~\ref{fig_graphs}B, this network retains the generally increasing 
dependence of \emph{II} upon $g_{\text{astro}}$ and $\lambda$, with the most 
notable difference being that \emph{II} is negative until $\lambda$ exceeds a 
certain threshold.

\begin{figure*} % [h!]
	A\includegraphics[width=0.95\columnwidth]{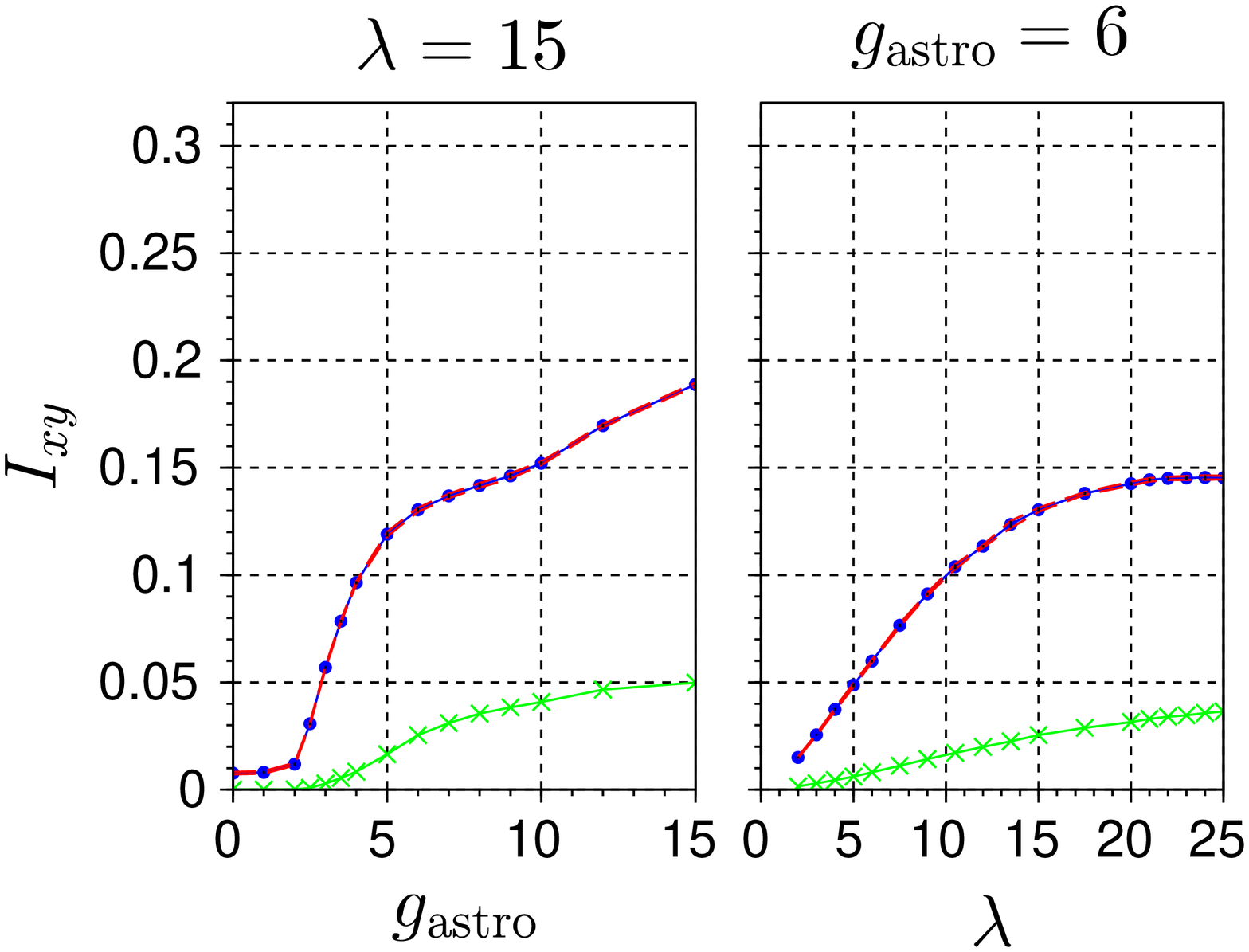}\hfill
	B\includegraphics[width=0.95\columnwidth]{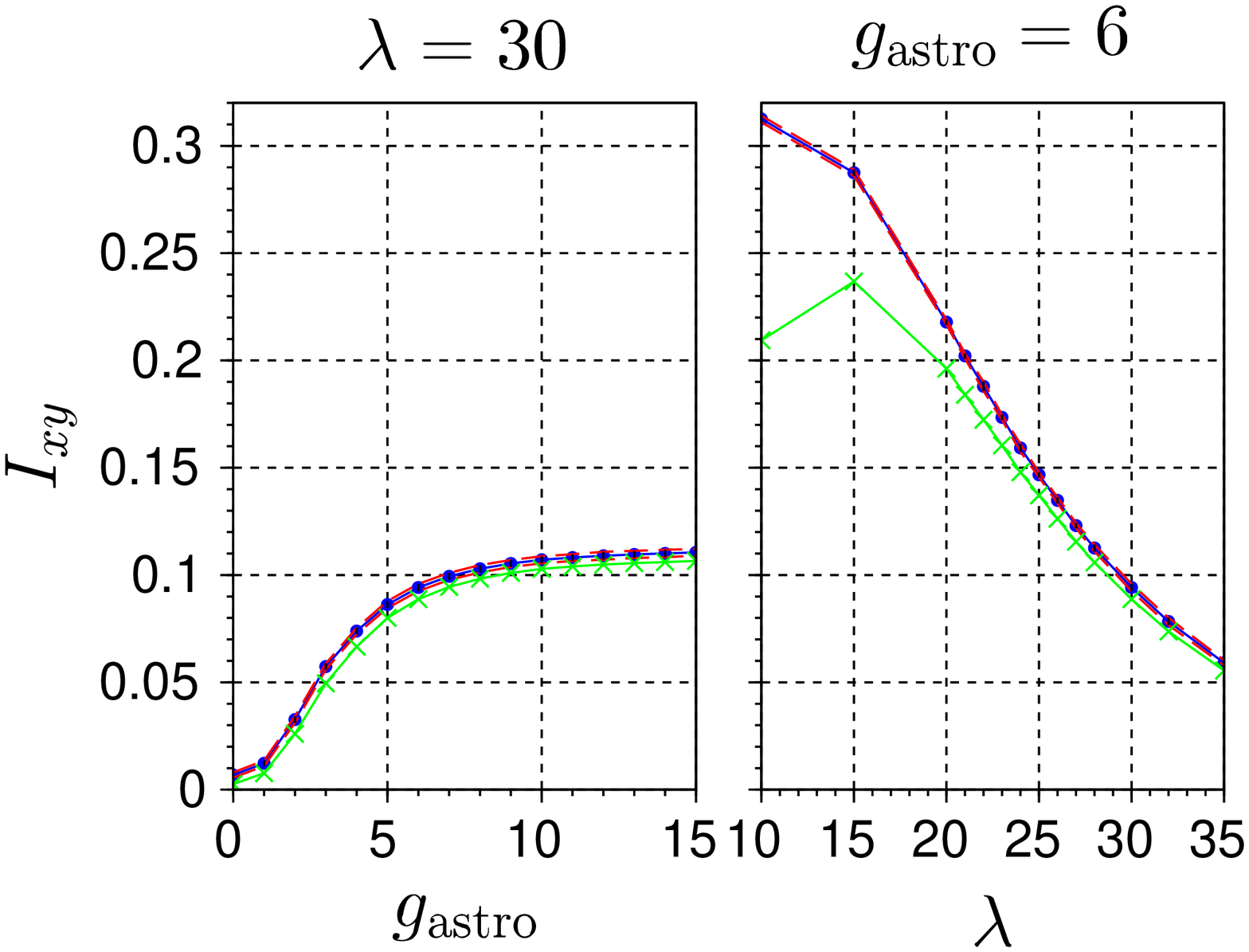}
	\caption{(Color online) Dependence of mutual information $I_{xy}$ upon 
neuro-astrocytic interaction $g_{\text{astro}}$ and upon average stimulation 
rate $\lambda$: A --- in random network (instance shown in 
Fig.~\ref{fig_topo}A); B --- in all-to-all network (Fig.~\ref{fig_topo}B). 
Legend as in Fig.~\ref{fig_graphs}.}\label{fig_Ixy}
\end{figure*}

To confirm the capacity of \emph{II} as a quantitative indicator for properties 
of complex dynamics in application to the system under study, we 
additionally consider graphs of mutual information $I_{xy}$ in the same 
settings, see Fig.~\ref{fig_Ixy} (note a greater range over $\lambda$ in 
Fig.~\ref{fig_Ixy}B as compared to Fig.~\ref{fig_graphs}B). Comparing 
Fig.~\ref{fig_graphs} to Fig.~\ref{fig_Ixy} we observe a qualitative difference 
in dependencies upon $\lambda$ in case of all-to-all network 
(Figs.~\ref{fig_graphs}B, \ref{fig_Ixy}B): while mutual information decreases 
with the increase of $\lambda$, \emph{II} is found to grow, and transits 
from negative to positive values before reaching its maximum. It means that 
even while the overall 
predictability of the system is waning, the system becomes more integrated in 
the sense that the advantage in this predictability when the system is taken as 
a whole over considering it by parts is found to grow. This confirms the 
capability of \emph{II} to capture features of complex dynamics that are not 
seen when using only mutual information.

Our analytical consideration is based upon mimicking the spiking-bursting 
dynamics by a model stochastic process which admits analytical calculation of 
effective information. We define this process $\xi(t)$ as a superposition of a 
time-correlated 
dichotomous component which turns system-wide bursting on and off, and a 
time-uncorrelated component describing spontaneous activity which occurs in the 
absence of a burst, in the following way.

At each instance of time the state of the dichotomous component can be either 
``bursting'' with probability $p_b$, or ``spontaneous'' (or
``spiking'') with probability $p_s=1-p_b$. While in the bursting mode,
the instantaneous state of the resulting process $x=\xi(t)$ is given by 
all ones: $x=11..1$ 
(further abbreviated as $x=1$). In case of spiking, the state $x$ is a random 
variate described by a discrete probability distribution $s_x$, so that the 
resulting one-time state probabilities read
\begin{subequations}\label{eq_Ponetime}
\begin{align}
p(x\neq1) &= p_s s_x,\\
p(x=1)    &= p_1, \quad p_1=p_s s_1 + p_b,\label{eq_Ponetime_p1}
\end{align}
\end{subequations}
where $s_1$ is the probability of spontaneous occurrence of $x=1$ in the 
absence of a burst (all neurons spontaneously spiking within the same time 
discretization window).

To describe two-time joint probabilities for $x=\xi(t)$ and $y=\xi(t+\tau)$, we 
consider a joint
state $xy$ which is a concatenation of bits in $x$ and $y$. The
spontaneous activity is assumed to be uncorrelated in time:
$s_{xy}=s_x s_y$. The time correlations of the dichotomous component are
described by a $2\times 2$ matrix of probabilities $p_{ss}$,
$p_{sb}$, $p_{bs}$, $p_{bb}$ which denote joint
probabilities to observe the respective spiking and/or bursting states in $x$
and $y$. The probabilities obey $p_{sb}=p_{bs}$ (due to
stationarity), $p_b=p_{bb}+p_{sb}$, $p_s=p_{ss}+p_{sb}$, thereby allowing to 
express all one- and two-time probabilities describing the dichotomous 
component in terms of two quantities, for which we chose $p_b$ and correlation 
coefficient $\phi$ defined by
\beq
p_{sb}=p_s p_b (1-\phi).
\eeq
The two-time joint probabilities for the resulting process are then expressed as
\begin{subequations}\label{eq_Ptwotime}
\begin{gather}
\begin{align}
p(x\neq 1, y\neq 1) &= p_{ss} s_x s_y, \\
p(x\neq 1, y=1) &= \pi s_x, \quad p(x=1, y\neq 1) = \pi s_y, \\
p(x=1, y=1) &= p_{11},
\end{align}\\
\pi=p_{ss} s_1 + p_{sb}, \quad p_{11}=p_{ss}
s_1^2 + 2 p_{sb} s_1 + p_{bb}.\label{eq_Ptwotime_p11}
\end{gather}
\end{subequations}

Note that the above notations can be applied to any subsystem
instead of the whole system (with the same dichotomous component, as it is 
system-wide anyway).

For this spiking-bursting process, the expression for
mutual information of $x$ and $y$ \eqref{eq_defIxy} after substitution of 
probabilities \eqref{eq_Ponetime}, \eqref{eq_Ptwotime} and algebraic 
simplifications reduces to
\begin{multline}\label{eq_Ixy}
I_{xy}=2(1-s_1)\{p_s\} + 2\{p_1\} -(1-s_1)^2\{p_{ss}\} -\\
-2(1-s_1)\{\pi\} - \{p_{11}\} = J(s_1;p_b, \phi),
\end{multline}
where we denote $\{q\}=-q \log_2 q$ for compactness.
With expressions for $p_1$, $p_{11}$,
$\pi$ from \eqref{eq_Ponetime_p1}, \eqref{eq_Ptwotime_p11} taken into account, 
$I_{xy}$ can be viewed as a function of $s_1$, denoted in \eqref{eq_Ixy} as 
$J(\cdot)$, with two parameters $p_b$ and $\phi$ characterizing the dichotomous 
(bursting) component.

\begin{figure}
	\centering
	\includegraphics[width=0.95\columnwidth]{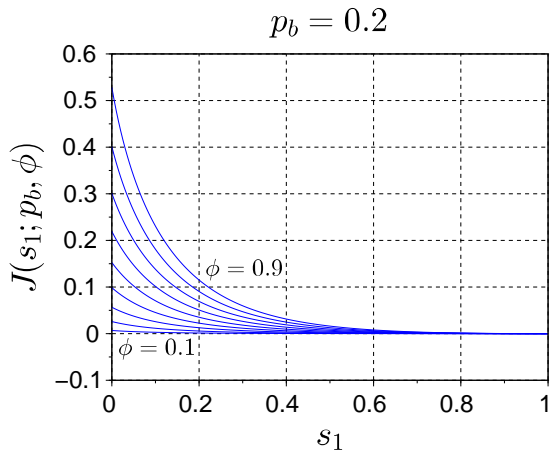}
	\caption{Family of plots of $J(s_1;p_b, \phi)$ at $p_b=0.2$ and $\phi$ 
varied from 0.1 to 0.9 with step 0.1.}\label{fig_Js1}
\end{figure}

A typical family of plots of $J(s_1;p_b, \phi)$ versus $s_1$ at $p_b=0.2$ and 
$\phi$ varied from 0.1 to 0.9 is shown in Fig.~\ref{fig_Js1}. Important 
particular cases are $$J(s_1=0) = 2\{p_s\} + 2\{p_b\} - \{p_{ss}\} - 
2\{p_{sb}\} - \{p_{bb}\}>0$$ which is the information of the dichotomous 
component alone; $J(s_1=1)=0$ (degenarate case --- ``always on'' deterministic 
state); $J(s_1)\equiv 0$ for any $s_1$ when $p_b=0$ or $\phi=0$ (absent or 
time-uncorrelated bursting). Otherwise, $J(s_1)$ is a positive decreasing 
function on $s_1\in[0,1)$.

Derivation of \eqref{eq_Ixy} does not impose any assumptions on the specific
type of the spiking probability distribution $s_x$. In particular,
spikes can be correlated across the system (but not in time).
Note that \eqref{eq_Ixy} is applicable as well
to any subsystem $A$ ($B$), with $s_1$ replaced by $s_A$ ($s_B$) which denotes 
the probability of a subsystem-wide simultaneous (within the same time 
discretization window) spike $x_A=1$ ($x_B=1$) in the absence of a burst, and
with same parameters of the dichotomous component (here $p_b$, $\phi$).
Effective information \eqref{eq_Ieff} is then written as
\beq\label{eq_IeffJ}
I_{\text{eff}}(AB)=J(s_1)-J(s_A)-J(s_B).
\eeq

Since as mentioned above $p_b=0$ or $\phi=0$ implies $J(s_1)=0$ for any $s_1$, 
this leads to $I_{\text{eff}}=0$ for any bipartition, and, accordingly, to zero 
\emph{II}, which agrees 
with our simulation results (left panels in Fig.~\ref{fig_graphs}A,B), where 
this case corresponds to the absence of coordinated activity induced by 
astrocytes ($g_{\text{astro}}=0$).

Consider the case of independent spiking with
\beq\label{eq_indep}
s_1=\prod_{i=1}^N P_i,
\eeq
where $P_i$ is the spontaneous spiking probability for an individual bit 
(neuron). Then $s_A=\prod_{i\in A} P_i$, $s_B=\prod_{i\in B} P_i$, $s_1=s_A 
s_B$. Denoting $s_A=s_1^{\nu}$, $s_B=s_1^{1-\nu}$, we rewrite \eqref{eq_IeffJ} 
as
\beq \label{eq_IeffNu}
I_{\text{eff}}(s_1; \nu)=J(s_1)-J(s_1^{\nu})-J(s_1^{1-\nu}),
\eeq
where $\nu$ is determined by the particular bipartition $AB$.

\begin{figure*}
	A\includegraphics[width=0.95\columnwidth]{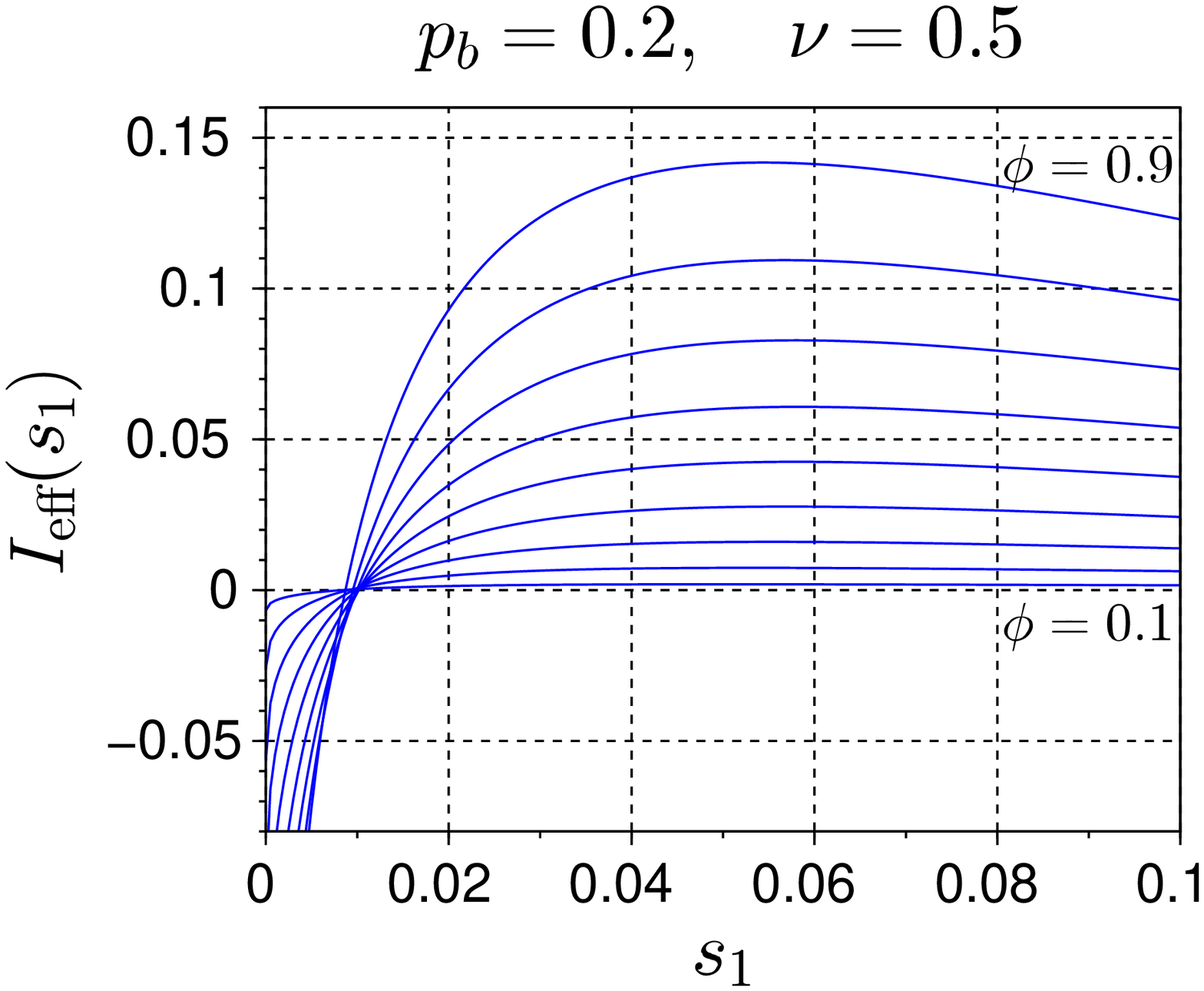}\hfill
	B\includegraphics[width=0.95\columnwidth]{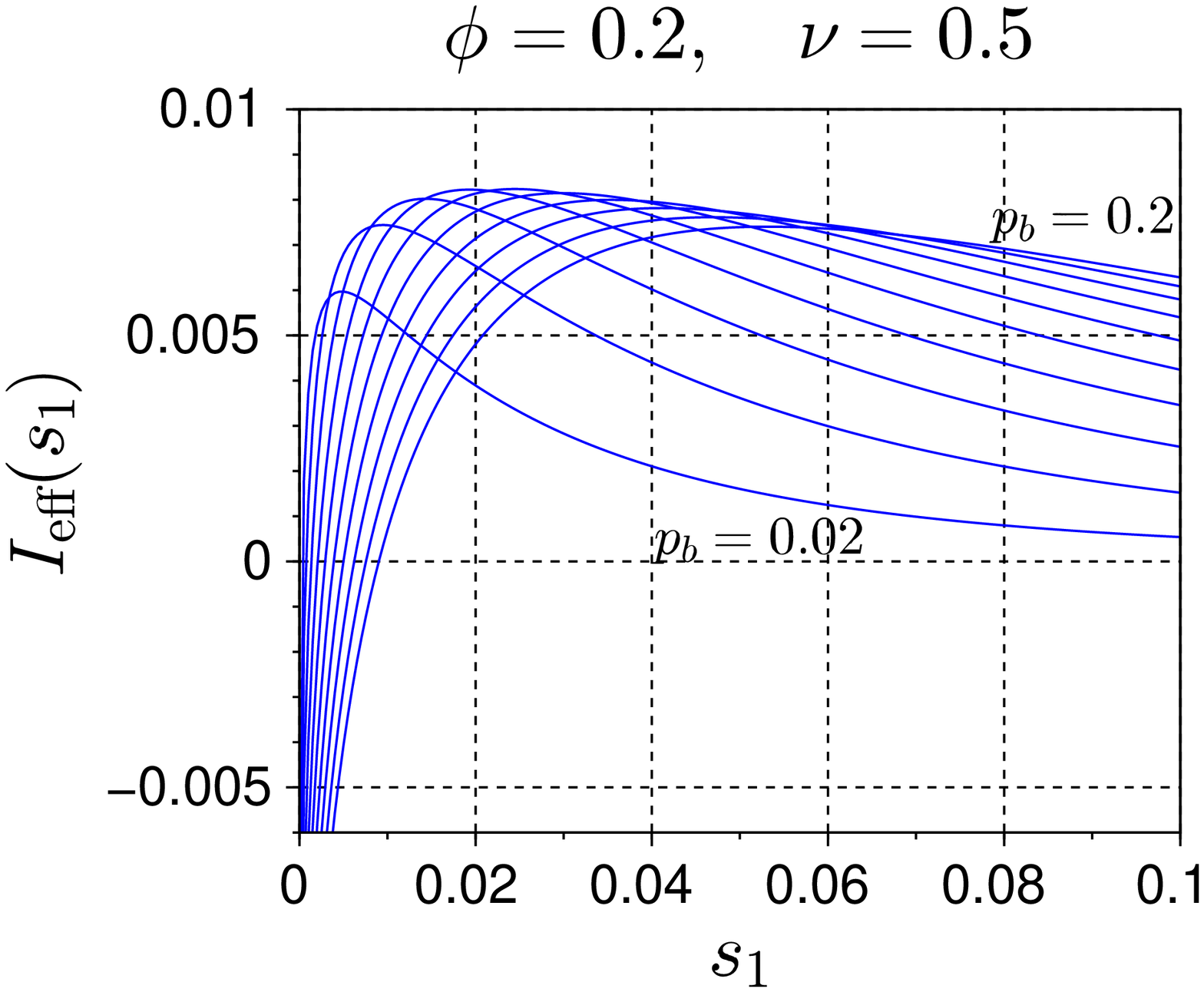}
	\caption{Families of plots of $I_{\text{eff}}(s_1;\nu=0.5)$: A --- at 
$p_b=0.2$ and $\phi$ varied from 0.1 to 0.9 with step 0.1; B --- at $\phi=0.2$ 
and $p_b$ varied from 0.02 to 0.2 with step 0.02.}\label{fig_Ieffs1}
\end{figure*}

Figure~\ref{fig_Ieffs1} shows typical families of plots of 
$I_{\text{eff}}(s_1;\nu=0.5)$ at $p_b=0.2$ and $\phi$ varied from 0.1 to 0.9  
in panel A (with increase of $\phi$, maximum of $I_{\text{eff}}(s_1)$ grows), 
and at $\phi=0.2$ with $p_b$ varied from 0.02 to 0.2 in panel B (with increase 
of $p_b$, root and maximum of $I_{\text{eff}}(s_1)$ shift to the right).

Hereinafter assuming $\phi\neq0$ and $p_b\neq0$, we notice the following: 
firstly, 
$I_{\text{eff}}(s_1=0)=-J(0)<0$, which implies $\mathit{II}<0$; secondly, 
$I_{\text{eff}}(s_1=1)=0$; thirdly, at $\phi>0$ function $I_{\text{eff}}(s_1)$ 
has a root and a positive maximum in interval $s_1\in(0,1)$. It implies that 
absent or insufficient spontaneous spiking activity leads to negative 
\emph{II}, while the increase in spiking turns \emph{II} positive. This is 
exactly observed in the all-to-all network simulation results, where spiking is 
determined by $\lambda$, see Fig.~\ref{fig_graphs}B (right panel). It can be 
additionally noticed in Fig.~\ref{fig_Ieffs1} that the root of 
$I_{\text{eff}}(s_1)$ (which is 
essentially the threshold in $s_1$ for positive \emph{II}) shows a stronger 
dependence upon the burst probability $p_b$ than upon correlation coefficient 
of bursting activity $\phi$.

Furthermore, expanding the last term of \eqref{eq_IeffNu} in powers of $\nu$ 
yields
\beq\label{eq_IeffExpand}
I_{\text{eff}} = -J(s_1^{\nu}) + \nu \cdot s_1 \log s_1 J'(s_1)+\dots \ .
\eeq
Consider the limit of large system $N\to \infty$ and a special bipartition with 
subsystem $A$ consisting of only one bit (neuron). Assuming that individual 
spontaneous spike probabilities of neurons $P_i$ in \eqref{eq_indep} retain 
their order of magnitude (in particular, do not tend to 0 or 1), we get
\beq
s_1\to +0, \quad s_1^{\nu}=s_A=O(1), \quad \nu \to +0,
\eeq
and finally $I_{\text{eff}}<0$ from \eqref{eq_IeffExpand}, which essentially 
prohibits positive \emph{II} in the spiking-bursting model for large systems.

The mentioned properties of $I_{\text{eff}}$ dependence upon parameters  can 
also be deduced from purely qualitative considerations in the sense of the 
reasoning in the end of Section~\ref{sec_meth_mod}. Absence of time-correlated 
bursting ($p_b=0$ or $\phi=0$), with only spiking present (which is 
time-uncorrelated), implies 
absence of predictability and thus zero \emph{II}. Absence of spontaneous 
spiking ($s_1=0$ in \eqref{eq_IeffNu}) implies complete synchronization (in 
terms of the binary process), and consequently highest overall 
predictability (mutual information), but negative \emph{II}. The presence of 
spontaneous activity decreases the predictability of the system as a whole, as 
well as that of any subsystem. According to \eqref{eq_Ieff}, favorable for 
positive $I_{\text{eff}}$ (and thus for positive \emph{II}) is the case when 
the predictability of subsystems is hindered more than that of the whole 
system. Hence the increasing dependence upon $s_1$: since in a system with 
independent spiking we have $s_1=s_A s_B < \min\{s_A, s_B\}$, spontaneous 
activity has indeed a greater impact upon  predictability for subsystems than 
for the whole system, thus leading to an increasing dependence of 
$I_{\text{eff}}$ upon $s_1$. This may eventually turn $I_{\text{eff}}$ positive 
for all bipartitions, which implies positive \emph{II}. 

\begin{figure*} % [h!]
	A\includegraphics[width=0.95\columnwidth]{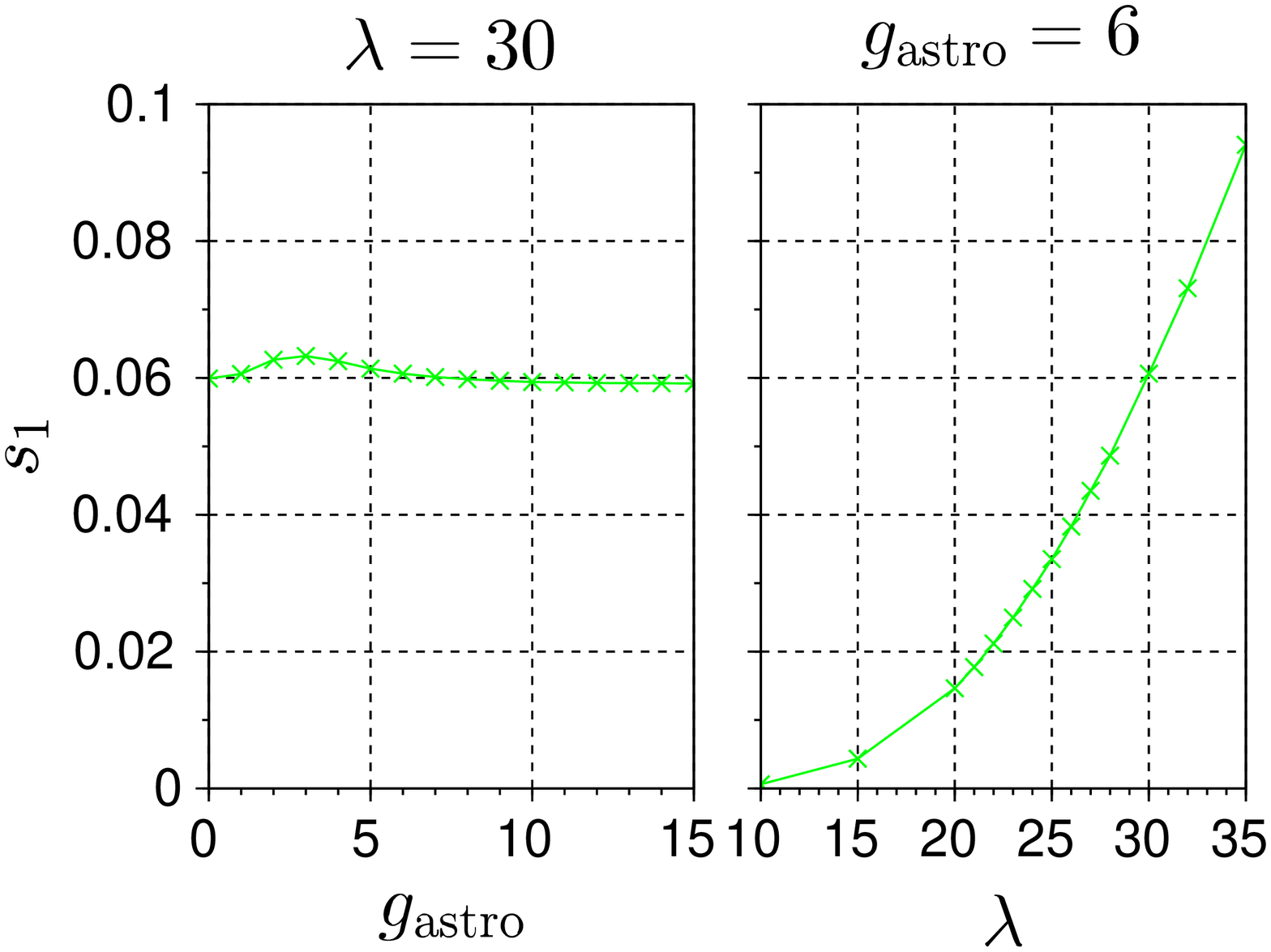}\hfill
	B\includegraphics[width=0.95\columnwidth]{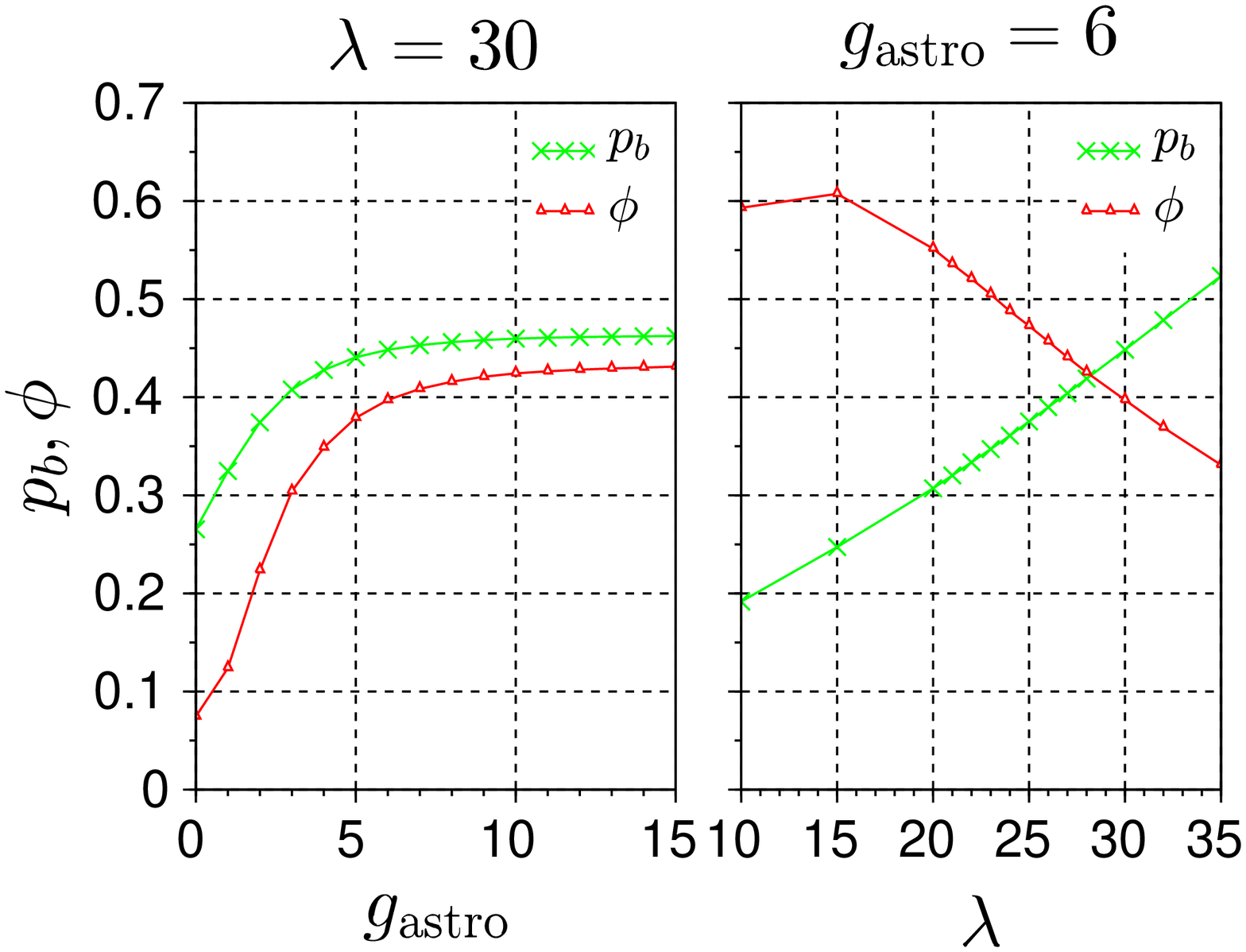}
	\caption{(Color online) Parameters of the spiking-bursting model $s_1$, 
$p_b$, $\phi$ fitted to simulated time series in case of all-to-all neuronal 
network.
	}\label{fig_pars}
\end{figure*}

In order to apply our analytical results to the networks under study, we fitted 
the parameters of the spiking-bursting process under the assumption of 
independent spiking \eqref{eq_indep} to the empirical probabilities from each 
simulation time series. The calculated values of $s_1$, $p_b$, $\phi$ in case 
of all-to-all neuronal network are plotted in Fig.~\ref{fig_pars} versus 
$g_{\text{astro}}$ and $\lambda$ (results for random network not shown due to 
an inferior adequacy of the model in this case, see below). As expected, 
spontaneous activity (here 
measured by $s_1$) increases with the rate of Poissonian stimulation $\lambda$ 
(Fig.~\ref{fig_pars}A, right panel), 
and time-correlated component becomes more pronounced (which is quantified by a 
saturated increase in $p_b$ and $\phi$) with the increase of astrocytic impact 
$g_{\text{astro}}$ (Fig.~\ref{fig_pars}B, left panel).

In Figs.~\ref{fig_graphs}, \ref{fig_Ixy} we plot the (semi-analytical) result 
of \eqref{eq_Ixy}, \eqref{eq_IeffJ} with the estimates substituted for $s_1$, 
$p_b$, $\phi$, and with bipartition $AB$ set to the actual minimum information 
bipartition found in the simulation. For the all-to-all network 
(Figs.~\ref{fig_graphs},\ref{fig_Ixy}B) this result is in good agreement with 
the direct calculation of $I_{xy}$ and \emph{II} (failing only in the region 
$\lambda<20$, see Fig.~\ref{fig_Ixy}B), unlike in case of random network 
(Figs.~\ref{fig_graphs},\ref{fig_Ixy}A), where the spiking-bursting model 
significantly underestimates both $I_{xy}$ and \emph{II}, in particular, giving 
negative values of \emph{II} where they are actually positive.

\section{Discussion}

We have demonstrated the generation of positive \emph{II} in neuro-astrocytic 
ensembles as a result of interplay between spontaneous (time-uncorrelated) 
spiking activity and astrocyte-induced coordinated dynamics of neurons. The 
analytic result for spiking-bursting stochastic model qualitatively and 
quantitatively reproduces the behavior of \emph{II} in the all-to-all network 
with all excitatory neurons (Fig.~\ref{fig_graphs}B). In particular, the 
existence of analytically predicted threshold in spontaneous activity for 
positive \emph{II} is observed.

Moreover, the 
spiking-bursting process introduced in this paper may be viewed as a simplistic 
but generic mechanism of generating positive \emph{II} in arbitrary ensembles. 
Complete analytic characterization of this mechanism is provided. In 
particular, it is shown that time correlated system-wide bursting and time 
uncorrelated spiking are both necessary ingredients for this mechanism to 
produce positive \emph{II}. Due to the simple and formal construction of the 
process, thus obtained positive \emph{II} must have no connection to 
consciousness in the underlying system, which may be seen as a counterexample 
to the original intent of \emph{II}. That said, it was also shown that 
\emph{II} of the spiking-bursting process is expected to turn negative when 
system size is increased. Aside from consciousness considerations, it means at 
least that positive \emph{II} in a large system requires a less trivial type 
of spatio-temporal patterns than one provided by the spiking-bursting model.

The increasing dependence of \emph{II} upon neuro-astrocytic interaction 
$g_{\text{astro}}$ and upon the intensity of spiking activity determined by 
$\lambda$ in a range of parameters is also observed in a more realistic random 
network model containing both excitatory and inhibitory synapses 
(Fig.~\ref{fig_graphs}A), for which our analysis is not directly applicable 
though. Remarkably, the decrease of $\lambda$ in the random network, in 
contrast to the all-to-all network, does not lead to negative \emph{II}. In 
this sense the less trivial dynamics of the random network appears to be even 
more favorable for positive \emph{II} than the spiking-bursting dynamics of the 
all-to-all network. This may be attributed to more complex astrocyte-induced 
space-time behavior, as compared to coordinated bursting alone, although we 
have not established specific connections of \emph{II} with properties of 
activation patterns in the random network. Nonetheless, based on this 
observation we also speculate that the limitation on network size which was 
predicted above for spiking-bursting dynamics may be lifted, thus allowing 
astrocyte-induced positive \emph{II} in large neuro-astrocytic networks. This 
is in line with the hypothesis that the presence of astrocytes may be crucial 
in producing complex collective dynamics in brain. The extension of our study 
to large systems is currently constrained by computational complexity of direct 
calculation of \emph{II} which grows exponentially with system size. Methods of 
entropy estimation by insufficient data 
\cite{archer2013bayesian,2017arXiv170802967T} may prove useful in this 
challenge, but will require specific validation for this task.

 \section*{Acknowledgments}
	This work was supported by the Russian Science Foundation Grant No. 
16-12-00077.

%\bibliographystyle{apsrev4-1}
%
%\bibliography{refs}


\end{document}